\documentclass[preprint,5p,number]{elsarticle} 

\usepackage{graphics}
\usepackage{amsmath}
\usepackage{float}
\usepackage{subfig}
\usepackage{tabularx}
\usepackage[normalem]{ulem}

\begin{document}
\title{Mechanisms of the refractive index change in DO11/PMMA due to photodegradation}
\author{Benjamin Anderson and  Mark G. Kuzyk}
\address{Department of Physics and Astronomy, Washington State University,
Pullman, WA 99164-2814}
\cortext[cor1]{Email: benjamin.r.anderson@gmail.com}
\date{\today}

\begin{abstract}
Using a white light interferometric microscope (WLIM) we measure the photodamage induced change in the complex index of refraction of disperse orange 11 (DO11) dye-doped (poly)methyl-methacrylate.  We find that the change in the imaginary part of the refractive index is consistent with previous measurements of photodamage-induced absorbance change.  Additionally, we find that the change in the real refractive index can be separated into a component due to damage to the dye molecules and a component due to irreversible damage to the polymer.

\vspace{1em}

\end{abstract}

\maketitle

\vspace{1em}

\section{Introduction}
Disperse orange 11 (DO11) is an organic laser dye \cite{howel02.01,howel04.01,Abbas09.01} which when doped into (poly)methyl-methacrylate (PMMA) is found to exhibit reversible photodegradation \cite{howel02.01,howel04.01,embaye08.01,ramini11.01,Anderson11.01,Ramini12.01,Ramini13.01,Anderson12.01,Anderson12.03,Anderson13.01,Anderson13.03,Anderson13.04}.  Previously, DO11/ PMMA's decay and recovery have been characterized using linear absorptive measurements (such as absorbance spectroscopy \cite{embaye08.01,ramini11.01,Anderson13.04} and transmittance imaging \cite{Anderson11.01, Anderson13.01, Anderson12.01,Anderson12.03,Anderson13.03, Anderson13.04}), nonlinear optical measurements  (specifically amplified spontaneous emission (ASE)\cite{howel02.01,howel04.01,embaye08.01,Ramini12.01}), and photo - conductivity\cite{Anderson12.01,Anderson12.03,Anderson13.03}. 

From these measurements a model of reversible photodegradation -- the extended correlated chromophore domain model (eCCDM) \cite{Ramini12.01,Ramini13.01,Anderson13.03, Anderson13.04} -- has been developed.  The eCCDM proposes that dye molecules form one-dimensional aggregates with the polymer host; where interactions between molecules result in a mitigation of photodegradation and recovery of damaged molecules.  While the exact mechanisms of the eCCDM are still under investigation, the model is found to predict all observed experimental data.

One of the fundamental questions still to be addressed arises from the observation of both a reversibly damaged and irreversibly damaged species in linear optical measurements \cite{Anderson11.01,Anderson11.02, Anderson12.01,Anderson12.03,Anderson13.01,Anderson13.03,Anderson13.04}, while amplified spontaneous emission (ASE) measurements observe full recovery \cite{howel02.01,howel04.01,embaye08.01,ramini11.01,Ramini12.01,Ramini13.01}.   As a possible explanation we proposed the hypothesis that the irreversibly damaged species is due to damage to the polymer host \cite{Anderson13.03,Anderson13.04}, as ASE probes only the dye molecules. If the irreversible species is due to polymer damage, we would expect the refractive index change to be the most sensitive probe as the polymer has the largest contribution to the sample's refractive index, given the typically low concentration of dye ($\approx 0.8$ wt \%).

\section{Method}
To measure the refractive index change of DO11/PMMA we use a white light interferometric microscope (WLIM) \cite{Anderson12.02,Anderson13.02} which uses white light in a Michelson interferometer and CCD camera to image the interferograms.  The WLIM allows for the resolution of a sample's complex index of refraction as a function of position in the plane of a thin film sample. The procedure to measure the change in refractive index using the WLIM is as follows.  Images, as a function of arm length difference, are first taken with the empty interferometer to measure the reference white light interferograms at each pixel.  Nearly identical samples are then placed in each of the interferometer's arms and the interferometer is adjusted to take into account the change in optical path length and phase distortion introduced by the samples.  Another set of images is then taken, as a function of arm-length difference, to measure the pristine sample's interferograms at each pixel.  After taking the white light and pristine interferograms, the sample in the stationary arm is burned using an ArKr laser operating at 488nm and focused to a line spot with a peak intensity of 5 W/cm$^2$.  The $x$-axis is defined to be the short axis, and the $y$-axis is the long axis.  The sample is photodegraded for 4 hours, at which point the pump beam is turned off and another set of images are taken to measure the damaged sample interferograms at each pixel.  After several days a final set of images are taken to determine changes due to recovery.



Once all the interferograms -- white light, pristine sample, damaged sample, and recovered sample -- are measured for each pixel, they are Fast Fourier Transformed to give spectral amplitude, $I(k_0;x,y)$, and complex phase, $\Phi(k_0;x,y)$ as a function of wavenumber and pixel position.  We then calculate the change in absorbance due to photodegradation (and recovery) using

\begin{align}
\Delta A&=A_2-A_1,
\\ &=-\ln \left(\frac{I_2}{I_0}\right)+\ln \left(\frac{I_1}{I_0}\right),
\\ &=-\ln \left(\frac{I_1}{I_2}\right),
\end{align}
where $A_2$ is the absorbance after degradation/recovery, $A_1$ is the pristine absorbance, $I_0$ is the spectral amplitude of the white light, $I_1$ is the spectral amplitude of the pristine sample, and $I_2$ is the spectral amplitude of the degraded sample. After determining the change in absorbance as a function of wavenumber and pixel position, we then calculate the change in phase:

\begin{equation}
\Delta\Phi(k_0;x,y)=\Phi_2(k_0;x,y)-\Phi_1(k_0;x,y),
\end{equation}
where $\Phi_1(k_0;x,y)$ is the pristine phase and $\Phi_2(k_0;x,y)$ is the phase after degradation/recovery.  

A complication arises when computing the change in phase as a function of position, as the process of unwrapping a noisy phase signal results in arbitrary constant offsets in the phase signal.  These offsets are found to be random from pixel-to-pixel with the average over a large number of pixels being zero.  However, using only a few pixels for averaging results in a non-zero phase offset.  Therefore, in order to accurately calculate the effect of photodegradation, we enforce a constraint that the phase change be zero at the wavenumber corresponding to peak absorbance change.  This constraint is consistent with the Kramers-Kronig relations between the real and imaginary parts of the refractive index\cite{Lucarini05.01,Anderson13.02}.  

After correcting for the random phase offset, the phase change is then converted to the scaled refractive index difference (SRID)\cite{Anderson13.02}:

\begin{align}
\xi&=\frac{\Phi_2-\Phi_1}{2k_0}
\\ &=d_1 \Delta n
\end{align}
where $d_1$ is the thickness of the damaged sample, and $\Delta n$ is the change in refractive index due to photodegradation (or recovery).

\section{Results and discussion}
As an example, consider the change in the complex index of refraction at nine points along the beam's $x$ axis at $x=$ 0, $\pm$ 12, $\pm$32, $\pm$41, and $\pm$46, where $x$ is in units of pixels.  Using the pump beam's known Gaussian profile we can convert each $x$ position into a pump beam intensity, and therefore the image data effectively measures the change in refractive index as a function of intensity.

\begin{figure}
\centering
\includegraphics{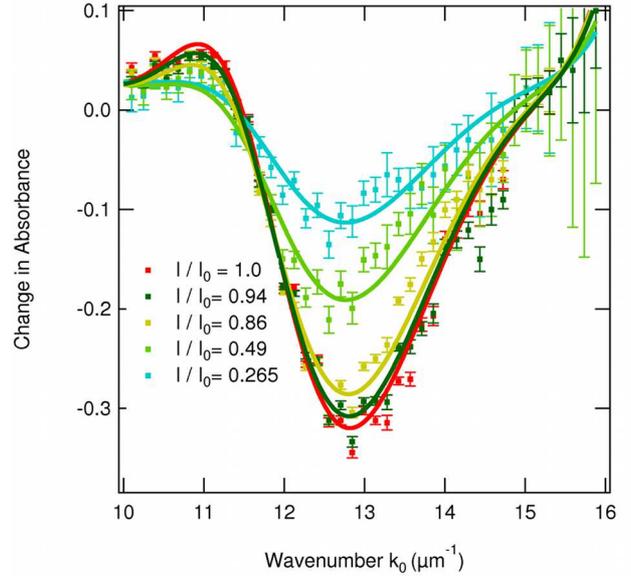}
\caption{Change in absorbance for several pump doses with three-peak Gaussian fits as a guide for the eye.}
\label{fig:dabs}
\end{figure}

\begin{figure}
\centering
\includegraphics{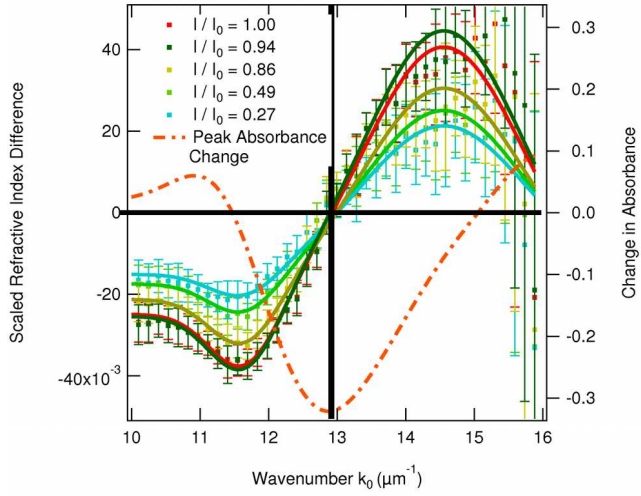}
\caption{Scaled refractive index difference for several doses with two-peak Gaussian fits for reference. Also shown is the peak change in absorbance.}
\label{fig:srid}
\end{figure}

Figure \ref{fig:dabs} shows the change in absorbance (points) as a function of wavenumber for five different intensities, with three peak Gaussian fits added as a guide for the eye.  The peak absorbance change and isosbestic points for the raw data are within experimental uncertainty of those measured previously using conventional spectrometers\cite{embaye08.01}.  In addition to confirming previous measurements of the change in absorbance, we also find the SRID as a function of wavenumber for five intensities as shown in Figure \ref{fig:srid}, where data is shown as points, and three-peak Gaussian fits are added as a guide for the eye.

\begin{figure}
\centering
\includegraphics{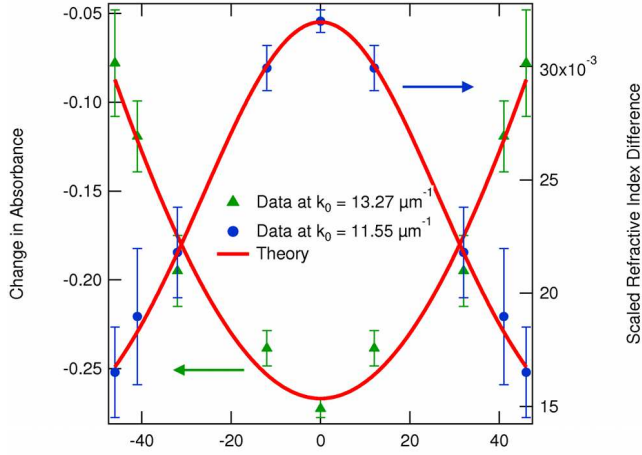}
\caption{Scaled refractive index change at $k_0=11.55$ $\mu$m$^{-1}$ (circles) and change in absorbance at $k_0=13.27$  $\mu$m$^{-1}$ (triangles) with Gaussian fits (curves).  The Gaussian widths of each fit are within experimental uncertainty of each other, and of the pump profile, as expected.}
\label{fig:space}
\end{figure}

To ensure the validity of our constraint on the isorefractive point -- that it occurs at the wavenumber of maximum absorbance change -- we consider the spatial profile of both the absorbance change and SRID. If our constraint is valid, the spatial profile of the absorbance change and SRID will be identical as both are proportional to the damage profile\cite{Anderson12.02,Anderson13.02}.  Choosing wavenumbers that correspond to maximum absorbance change and maximum SRID, we plot the absorbance change and SRID for each pixel in our data set and fit both curves to a Gaussian function as shown in Figure \ref{fig:space}.  Both data sets are found to agree with Gaussian fits with widths that are within experimental uncertainties.  This suggests that the constraint imposed on the isorefractive point is valid.

\begin{figure}
\centering
\includegraphics{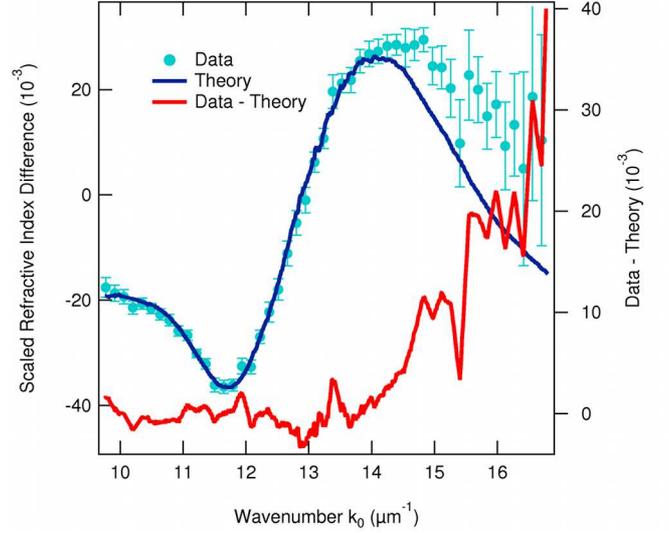}
\caption{Scaled refractive index change as measured by the WLIM at peak intensity (points), KK of absorbance data (blue curve), and residual between the two (red curve).}
\label{fig:diff}
\end{figure}

Given the success of our measurement of the SRID as a function of intensity, we also compute the SRID predicted by the Kramers-Kronig relations for comparison.   Figure \ref{fig:diff} shows the measured SRID at peak damage (points) to the Kramers-Kronig (KK) calculation (blue line) and the residual between data and theory (red line).  At longer wavelengths (smaller $k_0$) the KK calculation and the measured SRID are within experimental uncertainty of each other.  However, starting above $k_0$= 14 $\mu$m$^{-1}$ ($\lambda \approx $ 450nm) the measured SRID and KK calculation begin diverging.  Given that the difference is found to monotonically increase with wavenumber starting near $k_0$= 14$\mu$m$^{-1}$, we propose that this difference is due to irreversible photodegradation of the polymer host, as the measured difference correlates with measurements of the refractive index change of neat PMMA when photodegraded\cite{Baum10.01, Baum07.01, Watanabe12.01,Ahmed09.01,Murase99.01,Sowa06.01}.  

To understand why the KK calculation is insensitive to the contribution of polymer damage, we consider two limitations of the KK calculation.  First, the Hilbert transform integral spans a limited range of wavenumbers (corresponding to 390 nm $<\lambda <$ 800 nm). This wavenumber domain leaves out the spectral region in which the polymer's absorbance change is greatest ($\lambda <$ 350nm)\cite{Baum10.01, Baum07.01, Watanabe12.01,Ahmed09.01,Murase99.01,Sowa06.01}, but includes the region of peak absorbance change for DO11.  Therefore by not including the absorbance change in the UV regime, the KK calculation misses the effect of polymer damage.

\begin{figure}
\centering
\includegraphics{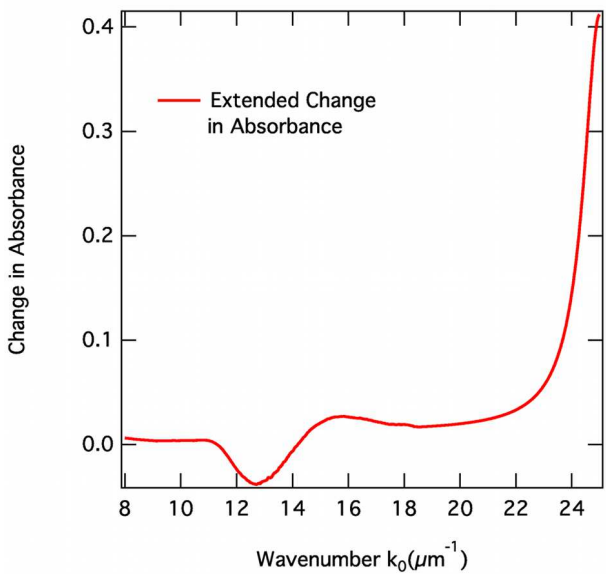}
\caption{Change in absorbance of DO11/PMMA with UV region approximated by the absorbance change of neat PMMA.}
\label{fig:absext}
\end{figure}

\begin{figure}
\centering
\includegraphics{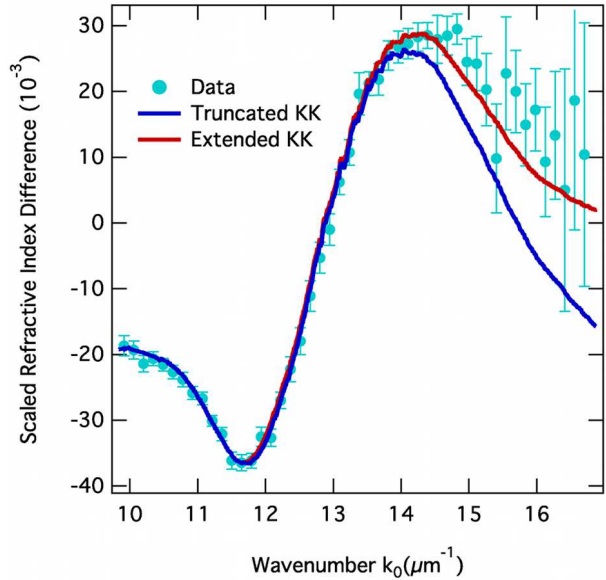}
\caption{Scaled refractive index change as measured by the WLIM at peak intensity as well as the truncated and extended KK calculation.  By including a peak in the absorbance change in the UV regime, we reproduce the measured SRID using the KK relations.}
\label{fig:kkext}
\end{figure}

As an estimate of the effect of irreversible polymer damage on the SRID, we approximate the change in absorbance in the UV regime, shown in Figure \ref{fig:absext}, and perform the KK calculation again.  We assume that the change in absorbance in the UV regime is of the same form as the change in absorbance for neat PMMA\cite{Baum10.01,Ahmed09.01,Baum07.01}. By performing the Hilbert transform on the extended change in absorbance we find the estimated SRID, as shown by the red curve in Figure \ref{fig:kkext}.  We find that by including a change in the UV absorption similar to that of neat PMMA our KK calculation is consistent within experimental uncertainty of the measured SRID.  This suggests that irreversible polymer damage is indeed responsible for the deviation between the truncated KK calculation and the measured SRID.

The second limitation to the KK calculation of the SRID is that the KK relations require the measured change in absorbance to be solely due to changes in the system's electric susceptibility.  However, irreversible photodegradation of PMMA is known to form scattering sites \cite{Watanabe12.01}, whose effect on absorption and refractive index does not follow the KK relations\cite{Cabannes21.01,Zimm54.01,Nakagaki56.01}.   The change in refractive index due to scattering sites is known to be proportional to $k_0^4$\cite{Cabannes21.01,Zimm54.01,Nakagaki56.01}, which is consistent with the observation of the divergence from the KK calculation. Though the absorption peak of PMMA is in the UV spectral range and thus requires an UV photon to cause and excitation and damage, absorption of a photon near the dopant's absorbance may lead to damage of the polymer through energy transfer\cite{Moshrefzadeh93.01,Gonzalez00.01,Chang01.01,Annieta,Rabek95.01}.

\section{Conclusions}
In conclusion, we spatially resolve the change in complex index of refraction due to photodegradation finding that the change in absorbance is consistent with previous measurements\cite{embaye08.01} while  the measured change in refractive index is inconsistent with predictions using KK relations.  Given the form of the residual, as shown in Figure \ref{fig:diff}, we conclude that the difference is due to irreversible polymer damage as the KK calculation primarily considers the effect of photodegradation on the dye molecules alone. This result supports the proposed nature of irreversible photodegradation in DO11/PMMA being due to polymer damage \cite{Anderson13.03,Anderson13.04}.   Further research is currently underway to test this hypothesis including UV spectroscopy, scattering experiments, and FTIR.

\section{Acknowledgments}
We would like to thank Elizabeth Bernhardt for help with sample preparation and would like to thank Wright Patterson Air Force Base and Air Force Office of Scientific Research (FA9550- 10-1-0286) for their continued support of this research.

\newpage

\bibliographystyle{model1-num-names}
\bibliography{PrimaryDatabase}

\end{document}